\documentstyle[12pt]{article}

\newcommand{\ba}{\begin{eqnarray}}
\newcommand{\ea}{\end{eqnarray}}
\input{psfig}
\def\ii{\'{\i}}
\begin{document}
\pagestyle{plain}

\title{Algebraic Approach to Baryon Structure}
\author{A. Leviatan\\
Racah Institute of Physics, The Hebrew University,\\
Jerusalem 91904, Israel
\and
R. Bijker\\
Instituto de Ciencias Nucleares, U.N.A.M., A.P. 70-543,\\
04510 M\'exico, D.F., M\'exico}
\date{}
\maketitle

\begin{abstract}
We present an algebraic approach to the internal structure of baryons
in terms of three constituents. We investigate a collective model 
in which the nucleon is regarded as a rotating and vibrating oblate 
top with a prescribed distribution of charges and magnetization.
We contrast the 
collective and single-particle descriptions of baryons
and compare the predictions of the model with existing data on masses, 
electromagnetic elastic and transition form factors and strong decays
widths.
\end{abstract}
\vspace{2cm}
\begin{center}
Invited talk at `Symmetries in Science IX',\\
Bregenz, Austria, August 6-10, 1996\\
\end{center}

\clearpage
\section{Introduction}
\setcounter{equation}{0}

In the usual description of baryons in terms of three constituents, the 
wave-function is a product of a space part and an internal
spin-flavor-color part. Algebraic methods have been used extensively in 
the past to describe the internal part in terms of 
the symmetry groups $SU_{sf}(6)\otimes SU_{c}(3)$ \cite{MGYN,GR}.
The difference between models lies in different assumptions on
the spatial dynamics of the three constituents. 
Quark potential models in nonrelativistic \cite{NRQM} or relativized
\cite{RQM} forms emphasize the single-particle aspects of quark dynamics 
for which only a few low-lying configurations in the confining potential 
contribute significantly to the eigenstates of the Hamiltonian. 
Excited baryons in this description correspond to single-particle 
excitations of levels in the confining potential. On the other hand,
flux-tube models, soliton models
as well as some regularities
in the observed spectra ({\it e.g.} linear Regge trajectories,
parity doubling) hint that an alternative, collective
type of dynamics may play a role in the structure of baryons.

The aim of the present contribution is to report on an algebraic
framework \cite{BIL} which encompasses both the single-particle  and 
collective scenarios. The formalism is based on a $U(7)$ spectrum generating
algebra whose bosonic realization and geometry were discussed at length in 
a previous contribution to these proceedings \cite{BL}. 
The algebraic formulation allows us to study
a large class of models, all with the same spin-flavor-color structure,
but different types of spatial dynamics.
Among the models is the familiar harmonic oscillator quark model
taken as a proto-type for single-particle dynamics,
and a collective model in which the baryon resonances
are interpreted as rotations and vibrations of an oblate-top
shaped string with a distribution of charge magnetic moments. In what 
follows we report on an application of the model to the mass
spectrum of nonstrange baryons, derive expressions of
form factors relevant for electromagnetic and strong couplings \cite{BIL}.
We examine the effect of spin-flavor breaking and stretching on the
electromagnetic helicity amplitudes \cite{emff} and calculate strong 
decay widths \cite{strong}.
The predictions of the collective-model are compared with those of the
single-particle description and with experimental data on these 
observables. The reader is encouraged to consult first \cite{BL} and 
to find more details in \cite{BIL,emff,strong,CONF}.
 
\section{Mass Spectrum}
\setcounter{equation}{0}

We consider the baryon as an object composed of three identical
constituents carrying internal quantum numbers: 
$\mbox{flavor}=\mbox{triplet}=u,d,s$;
$\mbox{spin}=\mbox{doublet}=1/2$; and $\mbox{color}=\mbox{triplet}$.
The geometric arrangement of this three-body system
can be phrased in
terms of two relative Jacobi vectors.
It is possible \cite{BIL,BL} 
to cast the dynamics of these spatial degrees of freedom
in algebraic form, in terms of a $U(7)$ spectrum generating algebra.
The full algebraic structure is obtained by combining
the geometric part, with the usual spin-flavor-color part,
resulting in total wave functions which are
representations of $U(7) \otimes SU_{sf}(6) \otimes SU_c(3)$,
\ba
\left| \, ^{2S+1}\mbox{dim}\{SU_f(3)\}_J \,
[\mbox{dim}\{SU_{sf}(6)\},L^P]_{(v_1,v_2);K} \, \right> ~. \label{wf}
\ea
The classification under
$SU_{sf}(6)$ and its flavor-spin subgroups $SU_f(3)\otimes SU_{S}(2)$,
are denoted by the dimension of the corresponding irreducible 
representations. We consider a collective model in which nonstrange
baryons are interpreted as excitations of the string configuration
in Fig.~1. In this case the spatial part in Eq.~(\ref{wf})
consists of an oblate-top wave function \cite{BL} characterized by
the labels: $(v_1,v_2);K,L^P_t$, where $(v_1,v_2)$
denotes the vibrations (stretching and bending) of the string; 
$K$ denotes the projection of the
rotational angular momentum $L$ on the body-fixed symmetry-axis; 
$P$ the parity and $t$ the $S_3$ symmetry type of the state under 
permutations. The permutation symmetry of the spatial part must be the
same as that of the spin-flavor part to ensure antisymmetry of the full
space-spin-flavor-color wave-function
($SU_{sf}(6)$ species: 
$[56]$ symmetric, $[20]$ antisymmetric, $[70]$ mixed symmetry).
The spin $S$ and $L$ are coupled to total angular momentum $J$.
In this notation the nucleon ($S=1/2$, flavor octet) and 
delta ($S=3/2$ flavor decuplet) ground state wave functions
are given by
\ba
\left| \, ^{2}8_{1/2} \, [56,0^+]_{(0,0);0} \, \right>
\hspace{1cm} \mbox{and} \hspace{1cm}
\left| \, ^{4}10_{3/2} \, [56,0^+]_{(0,0);0} \, \right> ~.
\label{ndelwf}
\ea
For harmonic oscillator dynamics, the spin-flavor part in 
is the same, while in the spatial part the $(v_1,v_2);K$ 
labels are replaced by the harmonic oscillator quantum number $n$.

An $S_3$-invariant operator can now be constructed along the lines outlined
in \cite{BL}, and used to describe properties of nucleon and delta
resonances. In the collective model the resulting mass spectrum exhibits
rotational states $(L^{P}_t,K)$ arranged in bands built on top of each 
vibration
$(v_1,v_2)$. 
For a large model space
the spectrum follows closely a mass formula of the form
\ba
M^2 = M^2_{0} + M^2_{\rm vib}(\lambda_1,\lambda_2) + 
      M^2_{\rm rot}(\alpha) + M^2_{\rm spin-flavor}(a,b,c) ~.
\ea
The spatial $U(7)$ contribution to the mass consists of a vibrational
part ($M^2_{\rm vib} = \lambda_1v_1+\lambda_2v_2$)
and a rotational part
($M^2_{\rm rot} = \alpha L$). The spin-flavor contribution is expressed
in a G\"ursey Radicati form \cite{GR} associated with the chain
$SU_{sf}(6)\supset SU_{f}(3)\otimes SU_s(2)$. 
The resulting fit (r.m.s deviation of 39 MeV)
for $3*$ and $4*$ nucleon and delta resonances
is shown in Fig.~2, along with a comparison to the non-relativistic
\cite{NRQM} and relativized \cite{RQM} quark models. As can be seen, 
the quality of the fits are comparable although the underlying dynamics 
and corresponding wave functions are different.
A typical oblate-top collective wave-function for the ground state is shown
in Fig.~3, expanded in an harmonic oscillator basis. 
It exhibits strong mixing of many oscillator shells and
reflects a correlated motion of the constituents.
This significant spread should be compared
to the structure of the ground state in the quark potential model
\cite{IKK} where $81 \%$ of the wave function is a pure $n=0$
configuration and the $19 \%$ admixture of $n=2$ components
is induced by the hyperfine interaction.
The above analysis of the spectrum in the nonstrange sector, 
shows that masses alone are not sufficient to distinguish between
single-particle and collective forms of dynamics. To do so require an
examination of other observables (e.g. electromagnetic and strong couplings)
which are more sensitive to the structure of wave-functions.

\section{Collective Form Factors}
\setcounter{equation}{0}

To consider decay processes of baryon resonances
we need two ingredients: (i) the wave function of the initial and final 
states and (ii) the form of the transition operator. 
It is assumed that the operators inducing the electromagnetic (strong)
transitions involve an absorption or emission of a photon 
(elementary meson) from a single constituent. In such 
circumstances, the couplings discussed below can be expressed 
in terms of the operators
\ba
\hat U &=& \mbox{e}^{ -ik \sqrt{\frac{2}{3}} \lambda_z} ~,
\nonumber\\
\hat T_{m} &=& \frac{i m_{3} k_0}{2} \left(
\sqrt{\frac{2}{3}} \, \lambda_m \,
\mbox{e}^{ -ik \sqrt{\frac{2}{3}} \lambda_z} \, + \,
\mbox{e}^{ -ik \sqrt{\frac{2}{3}} \lambda_z} \,
\sqrt{\frac{2}{3}} \, \lambda_m \right) ~, \label{ut}
\ea
where $\lambda_{m}$ ($m=0,\pm$) are Jacobi coordinates and 
$(k_0,\vec{k})$ is the four-momentum of the absorbed quanta.
The form factors of interest are proportional to the matrix elements of
these operators. In the algebraic approach, these operators are first  
mapped onto the $U(7)$ algebra
and their matrix elements are evaluated in the wave-functions of 
Eq.~(\ref{wf}).
The algebraic images of the operators in Eq.~(\ref{ut}) and
the calculation of matrix elements were presented in \cite{BL}. Table I
in \cite{BL} shows the resulting elementary form factors 
corresponding to the case in which the charge and magnetization are 
concentrated at the end points of the string of Fig.~1.
Different types of collective models are specified by
a distribution of the charge, magnetic moment, etc.,
along the string. For the present analysis
we use the (normalized) distribution
\ba
g(\beta) &=& \beta^2 \, \mbox{e}^{-\beta/a}/2a^3 ~, \label{gbeta}
\ea
where $\beta$ is a radial coordinate and $a$ is a scale parameter.
The collective form factors are obtained
by folding the matrix elements of $\hat U$ and
$\hat T_{m}$ with this probability distribution
\ba
{\cal F}(k)   &=& \int \mbox{d} \beta \, g(\beta) \,
\langle \psi_f | \hat U   | \psi_i \rangle ~,
\nonumber\\
{\cal G}_m(k) &=& \int \mbox{d} \beta \, g(\beta) \,
\langle \psi_f | \hat T_{m} | \psi_i \rangle  ~.
\label{radint}
\ea
Here $\psi$ denotes the spatial part of the baryon wave function.
The ansatz of Eq.~(\ref{gbeta}) for the probability distribution
is made to obtain the dipole form for the elastic form factor.
With the same distribution we can now derive closed expressions 
for inelastic form factors connecting other final states. A sample
of collective transition form factors of the
distributed string are given in Table~\ref{cff}, which also lists 
the corresponding harmonic oscillator form factors. 
The single-particle and collective form factors differ in their $k$ 
dependence. In particular, all collective form factors
drop as powers of $k$. This property is well-known experimentally and is 
in contrast with harmonic oscillator quark models in which all 
form factors which fall off exponentially.

\section{Electromagnetic Couplings}
\setcounter{equation}{0}

In electromagnetic processes such as photo- and electroproduction
we encounter both elastic transitions involving diagonal matrix elements 
and inelastic transitions involving non-diagonal matrix elements
of the transition operator. We consider below the corresponding observables
(elastic electric and magnetic form factors of the nucleon
and helicity amplitudes) which can be measured.

\subsection{Elastic Form Factors of the Nucleon}

Taking into account the overall symmetry of the nucleon wave function,
the elastic electric ($E$) and magnetic ($M$) collective form factors 
are given by
\ba
G^{N}_E &=& 3 \int \mbox{d} \beta \, g(\beta)\,
\langle \, \Psi ;\, M_J=1/2 \,| \,
e_3\, \hat U \, | \, \Psi ;\,
M_J=1/2 \, \rangle ~,
\nonumber\\
G^{N}_M &=& 3 \int \mbox{d} \beta \, g(\beta)\,
\langle \, \Psi ;\, M_J=1/2 \, | \,\mu_3\,
e_3\, \sigma_{3,z}\, \hat U \, | \, \Psi ;\,
M_J=1/2 \, \rangle ~,
\ea
where $\Psi$ denotes the nucleon wave function
$^28^N_{1/2}[56,0^+]_{(0,0);0}$ with $N=p\,(n)$ for proton (neutron).
Further $e_3$,
$\mu_3=eg_3/2m_{3}\,$, $m_3$, $g_3$, $s_3=\sigma_3/2$ are the charge
(in units of $e$: $e_u=2/3$, $e_d=-1/3$), scale magnetic moment, mass,
$g$-factor and spin, respectively, of the third constituent.
Using the results of Table~\ref{cff} we obtain
\ba
G_E^p &=& \frac{1}{(1+k^2a^2)^2}
\;\;\; ; \;\;\; 
G_E^n \; =\; 0 ~, 
\label{gepn}
\ea
for the charge form factors. The scale parameter in the distribution
(\ref{gbeta}) is related to the proton charge radius
$\langle r^2 \rangle_{E}^{p} = 12a^2$.
Similarly, for the magnetic form factors we obtain
\ba
G_M^p/\mu_p &=& G_M^n/\mu_n \; =\; \frac{1}{(1+k^2a^2)^2} ~,
\label{gmpn}
\ea
where the corresponding magnetic moments are
\ba
\mu_p &=& \mu \;\;\; ; \;\;\; \mu_n \; = \; -{2\over 3}\mu
\label{mupn}
\ea
respectively.
Here we have assumed spin-flavor symmetry so 
that the mass and the $g$-factor of
the up ($u$) and down ($d$) constituents are identical, $m_u=m_d=m_q$ and
$g_u=g_d=g$. Accordingly $\mu_u=\mu_d=\mu$ and $\mu = e\,g/2m_q$ in 
Eq.~(\ref{mupn}). 
The corresponding harmonic oscillator elastic form factors 
are obtained by replacing the dipole function $(1+k^2a^2)^{-2}$ 
by the exponential function $e^{-k^2\beta^2/6}$ with $\beta$ related 
to the harmonic oscillator size parameter.
The form factors in Eqs.~(\ref{gepn})-(\ref{gmpn}) (and their harmonic
oscillator analogues) satisfy $G_M^p = \mu G_E^p$ and the relations 
$G_E^n = 0$ and $G_M^n/G_M^p = -2/3$ for all values of
the momentum transfer. These relations are due to 
spin-flavor symmetry, but are not obeyed by the experimental data.
Within a truncated three-constituents configuration space,
to have a nonvanishing neutron electric form factor,
one must break $SU_{sf}(6)$ \cite{friar}. 
This breaking can be achieved in various ways, {\it e.g.}
by including in the mass operator a hyperfine interaction \cite{iks}, or
by distorting the oblate-top geometry, allowing
for a quark-diquark structure \cite{tzeng}.
Within the model discussed here 
we study the breaking of
the $SU_{sf}(6)$ symmetry by assuming a flavor-dependent distribution
of the charge and the magnetization along the strings of Fig.~1,
\ba
g_u(\beta) &=& \beta^2 \, \mbox{e}^{-\beta/a_u} /2a_u^3 ~,
\nonumber\\
g_d(\beta) &=& \beta^2 \, \mbox{e}^{-\beta/a_d} /2a_d^3 ~.
\label{gugd}
\ea
With this dependence and for small symmetry-breaking  
the electric nucleon form factors become
\ba
G_E^p &\approx& \frac{1}{(1+k^2\bar{a}^2)^2} \Bigl [\, 1 + {5\over 3}
          \frac{k^2\bar{a}^2}{(1+k^2\bar{a}^2)}\Delta\,\Bigr ] ~,
\nonumber\\
G_E^n &\approx& {4\over 3}\frac{k^2\bar{a}^2}{(1+k^2\bar{a}^2)^3}\Delta ~,
\label{elff}
\ea
to leading order in $\Delta$. Here 
$\Delta\equiv (a_d^2 - a_u^2)/\bar{a}^2$ with 
$\bar{a}^2 = (a_d^2 + a_u^2)/2$.

If the length of the string in Fig.~1 is slightly different for $u$ and $d$,
so is their mass and thus in principle, their magnetic moment.
Applying the same procedure to the magnetic form factors gives (to leading
order in $\Delta$)
\ba
G_M^p/\mu_p &\approx & \frac{1}{(1+k^2\bar{a}^2)^2} \Bigl \{\, 1 + 
          \frac{k^2\bar{a}^2}{(1+k^2\bar{a}^2)}
\Bigl [ {7\over 9} - {16\over 81}
\Bigl ({\mu_d - \mu_u\over \mu_p}\Bigr )\Bigr ]\Delta\,\Bigr \} ~,
\nonumber\\
G_M^n/\mu_n &\approx & \frac{1}{(1+k^2\bar{a}^2)^2} \Bigl \{\, 1 - 
          \frac{k^2\bar{a}^2}{(1+k^2\bar{a}^2)}
\Bigl [ {1\over 3} - {8\over 27}
\Bigl ({\mu_d - \mu_u\over \mu_n}\Bigr )\Bigr ]\Delta\,\Bigr \} ~.
\label{magff}
\ea
Here $\mu_u e_u$ and $\mu_d e_d$ are the magnetic moments of the $u$ and
$d$ constituents which determine the proton and neutron magnetic moments
\ba
\mu_p &=& (4\mu_u e_u - \mu_d e_d)/3 \;\;\; ; \;\;\;
\mu_n \; =\; (4\mu_d e_d - \mu_u e_u)/3 ~.
\ea

In the calculations reported below we have used the following procedure
to determine the parameters.
In all cases we take $g_u=g_d=1$.
We assume that the constituent masses
$m_u$ and $m_d$ are determined from the measured magnetic moments.
This fixes the scale magnetic moments $\mu_u$ and $\mu_d$.
The scale parameters $a_u$ and $a_d$ in the distributions (\ref{gugd})
are determined from a simultaneous fit to the
proton and neutron charge radii, and to the proton and neutron electric 
and magnetic form factors. For the calculations in which the $SU_{sf}(6)$ 
symmetry is satisfied ($\mu_u=\mu_d=\mu$ and $a_u=a_d=a$) this procedure
yields $a=0.232$ fm and
$\mu=\mu_p= 2.793$ $\mu_N$ ($=0.126$ GeV$^{-1}$),
which corresponds to a constituent mass of $m_u=m_d=0.336$ GeV.
For the calculations in which the $SU_{sf}(6)$ symmetry is broken
we find $a_u=0.230$ fm and $a_d=0.257$ fm (implying $\Delta=0.2$ in 
Eqs.~(\ref{elff})-(\ref{magff}) ), $\mu_u=2.777\mu_{N}$ 
($= 0.126$ GeV$^{-1}$), $\mu_d=2.915\mu_{N}$, ($=0.133$ GeV$^{-1}$)
corresponding to $m_u=0.338$ GeV and $m_d=0.322$ GeV, respectively. 

Fig.~4 shows the elastic electric
form factors of the proton and the neutron divided by the dipole form,
$F_D = 1/(1+Q^2/0.71)^2$. The division by $F_D$ emphasizes the
effect of the breaking of spin-flavor symmetry. 
Fig.~5 shows the results for the elastic magnetic form factors. 
We see that while the breaking of spin-flavor symmetry can
account for the non-zero value of $G_E^n$ and gives a good description
of the data, it worsens the fit to the proton electric and neutron
magnetic form factors.
This implies either that 
the simple mechanism for spin-flavor breaking of Eq.~(\ref{gugd}) does 
not produce the right phenomenology and
other contributions, such as polarization of the neutron into
$p+\pi^{-}$, play an important role in the neutron electric form factor
\cite{cmt}. (A coupling to the meson
cloud through $\rho$, $\omega$ and $\phi$ mesons is indeed expected to
contribute in this range of $Q^2$ \cite{vmd}.)
This conclusion (i.e. worsening the proton form factors)
applies also to the other mechanisms of spin-flavor symmetry
breaking mentioned above, such as that induced by a hyperfine interaction
\cite{iks} which gives $a_u < a_d$ (`moves the up quark to the center and
the down quark to the periphery'). This pattern is a
consequence of the fact that within the framework of constituent models
$G_E^p$, $G_E^n$, $G^p_M$ and $G_M^n$ are intertwined
as is clearly evident from the expressions in 
Eqs.~(\ref{elff})-(\ref{magff}).

\subsection{Helicity Amplitudes}

Other (observable) quantities of interest are the helicity amplitudes
in photo- and electroproduction. 
The transverse helicity amplitudes
between the initial (ground) state of the nucleon and the final
(excited) state of a baryon resonance are expressed as \cite{BIL}
\ba
A^{N}_{\nu} &=& 6 \sqrt{\frac{\pi}{k_0}} \, \Bigl [ \, k
\langle L,0;S,\nu   | J,\nu \rangle \, {\cal B} -
\langle L,1;S,\nu-1 | J,\nu \rangle \, {\cal A} \, \Bigr ] ~,
\label{helamp}
\ea
where $\nu=1/2$, $3/2$ indicates the helicity.
The orbit- and spin-flip amplitudes (${\cal A}$ and ${\cal B}$,
respectively) are given by
\ba
{\cal B} &=& \int \mbox{d} \beta \, g(\beta)\,
\langle \Psi_f;M_J=\nu | \,\mu_3\,
e_3\, s_{3,+}\, \hat U \, | \Psi_i;M_J^{\prime}=\nu-1 \rangle ~,
\nonumber\\
{\cal A} &=& \int \mbox{d} \beta \, g(\beta)\,
 \langle \Psi_f;M_J=\nu | \, \mu_3\,
e_3\, \hat T_{+}/g_3 \, | \Psi_i;M_J^{\prime}=\nu-1 \rangle ~.
\label{ab}
\ea
These observables correspond to an absorption process 
($\gamma + B^{\prime} \rightarrow B$) of a right-handed photon with
four-momentum ($k_0,\vec{k}=k\hat{z}$). In Eq.~(\ref{ab}) 
$| \Psi_i \rangle$ denotes the (space-spin-flavor) wave function
of the initial nucleon ($B^{\prime}$) with
$^{2}8^N_{1/2}[56,0^+]_{(0,0);0}$ and $N=p,n$, and, similarly,
$| \Psi_f \rangle$ that of the final baryon resonance ($B$).
In general, the ${\cal B}$ and ${\cal A}$ amplitudes
of Eq.~(\ref{ab}) are proportional to the collective form factors
${\cal F}$ and ${\cal G}_+$ of Eq.~(\ref{radint}), respectively.
The breaking of spin-flavor symmetry has also influence on the helicity
amplitudes of Eq.~(\ref{helamp}). They would
now be given in terms of flavor-dependent collective form factors
${\cal F}_u(k)$, ${\cal G}_{u,+}(k)$ and ${\cal F}_d(k)$,
${\cal G}_{d,+}(k)$, which depend on the size parameters,
$a_u$ and $a_d$, respectively. 

When comparing with the experimental data
one must still choose a reference frame which determines the relation
between the three-momentum $k^2$ and the four-momentum
$Q^2=k^2 - k_{0}^2$.
It is convenient to choose the equal momentum or Breit frame where
\ba
k^2 &=& Q^2 + \frac{(W^2-M^2)^2}{2(M^2+W^2)+Q^2} ~.
\ea
Here $M$ is the nucleon mass, $W$ is the mass of the resonance,
and $-Q^2$ can be interpreted as the mass squared of
the virtual photon (for elastic scattering we have $k^2 = Q^2$).
In assessing the quality of the fits it should be noted that there are
no adjustable parameters involved in the calculation
of helicity amplitudes. All the parameters
appearing in Eq.~(\ref{ab}) and in the distributions (\ref{gugd}) 
are those extracted previously from the elastic form factors.

Representative results for
transverse helicity amplitudes $A_{1/2}$
and $A_{3/2}$ are shown in Fig.~6 for the nucleon resonances
$N(1520)D_{13}$ and $N(1680)F_{15}$.
From this figure it is seen that the effect 
of spin-flavor breaking is rather small.
Only in those cases in which the amplitude with $SU_{sf}(6)$
symmetry is zero, the effect is of some relevance.
Such is the case with proton helicity amplitudes for the
$^{4}8_{J}[70,L^{P}]$ multiplet
(e.g. the $L^{P}=1^{-}$ resonances $N(1675)D_{15}$ and $N(1700)D_{13}$)
and with neutron helicity-3/2
amplitudes for the $^{2}8_{J}[56,L^{P}]$ multiplet
(e.g. the $L^{P}=2^{+}$ resonance $N(1680)F_{15}$).
The conclusion that one can draw from this analysis is that, 
for all purposes, with the exception of the electric 
form factor of the neutron, the breaking of spin-flavor symmetry
according to the mechanism of Eq.~(\ref{gugd}) is of little importance.

In a string-like model of hadrons one expects on the basis of QCD
\cite{johnsbars} that strings will elongate (hadrons swell)
as their energy increases. This effect can be
easily included in the present analysis by making the scale parameters
of the strings energy dependent. We use here the simple ansatz
\ba
a &=& a_0\Bigl ( 1 + \xi\,{W-M\over M}\Bigr ) ~, \label{stretch}
\ea
where $M$ is the nucleon mass and $W$ the resonance mass. This ansatz
introduces a new parameter ($\xi$), the stretchability of the string.

Figs.~7 shows the effect
of stretching on the helicity amplitudes for 
$N(1520)D_{13}$ and 
$N(1680)F_{15}$. It is seen that the effect of
stretching, especially if one takes the value $\xi\approx 1$ suggested
by the arguments of \cite{johnsbars} and the Regge behavior of nucleon
resonances (see {\it e.g.} Fig.~5 of \cite{BIL}), is rather large.
In particular, the data for $N(1680)F_{15}$ and $N(1520)D_{13}$ 
show a clear indication that the form
factors are dropping faster than expected on the basis of the dipole
form. (Of course for the elastic form actors there is no stretching.)
We suggest that future data at CEBAF and MAMI be used to analyze
the effects of stretching on the helicity amplitudes.

\section{Strong Couplings}
\setcounter{equation}{0}
\label{sec2}

We consider in this Section strong decays of baryons of the form
$B \rightarrow B^{\prime} + M$ \cite{strong}.
The process involves an emission (by one of the constituents in
$B$) of an elementary pseudoscalar meson 
meson ($M=\pi$ or $\eta$) with energy
$k_0=E_M=E_B-E_{B^{\prime}}$ and momentum
$\vec{k}=\vec{P}_M=\vec{P}-\vec{P}^{\prime}=k \hat z$. 
Here $\vec{P}=P_z \hat z$
and $\vec{P}^{\prime}$ ($=P^{\prime}_z \hat z$)
are the momenta of the initial ($B$) and final baryon ($B^{\prime}$).
The calculations are performed in the rest frame of $B$ ($P_z=0$).

In the collective model the strong couplings (similar to the 
electromagnetic couplings) are obtained by 
folding the matrix elements of the transition operator inducing the decay
with the distribution function $g(\beta)$ of Eq.~(\ref{gbeta}).
These collective matrix elements can be expressed in terms of
helicity amplitudes. For decays in which the initial baryon has angular
momentum $\vec{J}=\vec{L}+\vec{S}$ and in which the final baryon is either
the nucleon or the delta with wave functions (\ref{ndelwf}),
the (strong) helicity amplitudes are
\ba
A_{\nu}(k) 
&=&  \frac{1}{(2\pi)^{3/2} (2k_0)^{1/2}} \left[
\langle L, 0,S,\nu   | J,\nu \rangle \, \zeta_0 Z_0(k) + \frac{1}{2}
\langle L, 1,S,\nu-1 | J,\nu \rangle \, \zeta_+ Z_-(k) \right.
\nonumber\\
&& + \left. \frac{1}{2}
\langle L,-1,S,\nu+1 | J,\nu \rangle \, \zeta_- Z_+(k) \right] ~.
\label{anu}
\ea
The coefficients $\zeta_m$ are spin-flavor matrix elements \cite{strong} 
and $Z_m(k)$ ($m=0,\pm$) are the radial matrix elements
\ba
Z_0(k) 
&=& 6 \, [gk - \frac{1}{6}hk] \, {\cal F}(k)^{*}
- 6h \, {\cal G}_z(k)^{*} ~,
\nonumber\\
Z_{\pm}(k) 
&=& -6h \, {\cal G}_{\mp}(k)^{*} ~. \label{radme}
\ea
The coefficients $g$ and $h$ denote the strength of two terms
in the transition operator. The radial matrix elements $Z_{m}(k)$
involve the same collective form factors ${\cal F}(k)$, 
${\cal G}_{\mp}(k)$ discussed in Section (3).
The reason for the complex conjugation in Eq.~(\ref{radme})
is that here we consider an emission process, whereas 
Table~\ref{cff} shows the form factors for an absorption process. 

The decay widths for a specific channel are given by \cite{LeYaouanc}
\ba
\Gamma(B \rightarrow B^{\prime} + M) &=& 2 \pi \rho_f \,
\frac{2}{2J+1} \sum_{\nu>0} | A_{\nu}(k) |^2
\label{dw}
\ea
where $\rho_f$ is a phase space factor.
For all resonances with the same value of $(v_1,v_2),L^P$
this expression can be rewritten
in a more transparent form in terms of only two elementary partial
wave amplitudes $W_l(k)$,
\ba
\Gamma(B \rightarrow B^{\prime} + M)
\;=\; 2 \pi \rho_f \, \frac{1}{(2\pi)^3 2k_0} \,
\sum_{l=L \pm 1} c_l \left| W_l(k) \right|^2 ~. \label{width}
\ea
Here $l$ is the relative orbital angular momentum between the final
baryon and the emitted meson. It takes the values $l=L \pm 1$ (the
value $l=L$ is not allowed because of parity conservation).
For this set of resonances, the $k$ dependence 
is contained in the partial wave amplitudes $W_l(k)$, while the
dependence on the individual baryon resonance is contained in the
coefficients $c_l$. In the algebraic method,
the $W_l(k)$ can be obtained in closed form. For example,
the corresponding $S$ and $D$ elementary partial wave amplitudes are
\ba
W_0(k) &=& i \, \left\{
[gk-\frac{1}{6}hk] \frac{ka}{(1+k^2a^2)^2}
+h \, m_3k_0a \frac{3-k^2a^2}{(1+k^2a^2)^3} \right\} ~,
\nonumber\\
W_2(k) &=& i \, \left\{
[gk-\frac{1}{6}hk] \frac{ka}{(1+k^2a^2)^2}
-h \, m_3k_0a \frac{4k^2a^2}{(1+k^2a^2)^3} \right\} ~. \label{pw02}
\ea

Partial widths for other models of the nucleon and its resonances
can be obtained by introducing the corresponding expressions for the
elementary amplitudes $W_l(k)$. For example, the relevant
expressions in the harmonic oscillator quark model are
\ba
W_0(k) &=& \frac{i}{3} \left\{ [gk-\frac{1}{6}hk] k \beta
+ h m_3 k_0 \beta (3-\frac{k^2 \beta^2}{3}) \right\}
\mbox{e}^{-k^2 \beta^2/6} ~,
\nonumber\\
W_2(k) &=& \frac{i}{3} \left\{ [gk-\frac{1}{6}hk] k \beta
- \frac{1}{3} h m_3 k_0 \beta k^2 \beta^2 \right\}
\mbox{e}^{-k^2 \beta^2/6} ~,
\ea

Use of Eqs.~(\ref{width})-(\ref{pw02}) allows us to do a straightforward
and systematic analysis of the experimental data.
The calculations are performed in the rest frame of the decaying resonance, 
in which the relativistic expression for the phase space factor
$\rho_f$ as well as for the momentum $k$ of the emitted meson are retained.
The expressions for $k$ and $\rho_f$ are
\ba
k^2 &=& -m_M^2 + \frac{(m_B^2-m_{B^{\prime}}^2+m_M^2)^2}{4m_B^2} ~,
\nonumber\\
\rho_f 
&=& 4 \pi \frac{E_{B^{\prime}}(k) E_M(k) k}{m_B}
\ea
with $E_{B^{\prime}}(k)=\sqrt{m_{B^{\prime}}^2+k^2}$ and
$E_{M}(k)=\sqrt{m_{M}^2+k^2}$.

We consider here decays with emission of $\pi$ and $\eta$.
The calculated values depend on the
two parameters $g$ and $h$ in Eq.~(\ref{radme})
and on the scale parameter $a$ of Eq.~(\ref{gbeta}).
In the present analysis we determine these parameters
from a least square fit to the $N \pi$
partial widths (which are relatively well known) with the exclusion
of the $S_{11}$ resonances. For the latter the situation is not clear due
to possible mixing of $N(1535)S_{11}$ and $N(1650)S_{11}$ and
the possible existence of a third $S_{11}$ resonance \cite{LW}. As a
result we find $g=1.164$ GeV$^{-1}$ and $h=-0.094$ GeV$^{-1}$. 
The relative sign is consistent with a previous analysis of the strong
decay of mesons \cite{GIK} and with a derivation from the axial-vector
coupling (see {\it e.g.} \cite{LeYaouanc}).
The scale parameter, $a=0.232$ fm, extracted in the present fit
is found to be equal to the value extracted in the calculation of
electromagnetic couplings \cite{emff}.
We keep $g$, $h$ and $a$ equal
for {\em all} resonances and {\em all} decay channels ($N \pi$,
$N \eta$, $\Delta \pi$, $\Delta \eta$). 
In comparing with previous calculations, it should be noted that in the 
calculation in the nonrelativistic quark model of \cite{KI}
the decay widths are parametrized by four reduced partial wave amplitudes
instead of the two elementary amplitudes $g$ and $h$. Furthermore,
the momentum dependence of these reduced amplitudes are represented
by constants. The calculation in the relativized quark model of \cite{CR}
was done using a pair-creation model for the decay and
involved a different assumption on the phase space factor.
Both the nonrelativistic and relativized quark model calculations
include the effects of mixing induced by the hyperfine interaction,
which in the present calculation are not taken into account.

The calculations of decay widths of 3* and 4* resonances
into the $N \pi$ and $\Delta\pi$ channels
are in fair agreement with experiment. 
Representative results are shown in Fig.~8.
The results are to a large
extent a consequence of spin-flavor symmetry. 
The use of `collective'
form factors improves somewhat the results when compared with older
(harmonic oscillator) calculations. This is shown in Table~\ref{regge}
where the decay of a $\Delta$ Regge trajectory into $N \pi$
is analyzed and compared with the calculations of \cite{LeYaouanc},
which are based on the harmonic oscillator model discussed in \cite{Dalitz}.
We also include the results of more recent calculations in the
nonrelativistic quark model \cite{KI} and in the relativized quark model
\cite{CR}. There does not seem to be anything unusual in the decays into
$\pi$ and our analysis confirms the results of previous analyses.

Contrary to the decays into $\pi$,
the decay widths into $\eta$ have some unusual
properties. The calculation gives systematically small values
for these widths. This is due to a combination of phase space
factors and the structure of the transition operator. Both depend
on the momentum transfer $k$, however,
due to the difference between
the $\pi$ and $\eta$ mass, the momentum carried by the $\eta$ is
smaller than that carried by the $\pi$. Therefore, the $\eta$ decay widths
are suppressed relative to the $\pi$ decays. The spin-flavor part is
approximately the same for $N \pi$ and $N \eta$, since $\pi$ and $\eta$
are in the same $SU_f(3)$ multiplet. We emphasize here, that
the transition operator was determined
by fitting the coefficients $g$ and $h$ to the $N \pi$ decays
of the 3* and 4* resonances. Hence the $\eta$ decays are
calculated without introducing any further parameters.

The experimental situation is unclear. The 1992 Particle Data Group (PDG)
compilation gave systematically small widths ($\sim 1$ MeV) for all
resonances except $N(1535)S_{11}$. The 1994 PDG compilation
deleted all $\eta$ widths with the exception of $N(1535)S_{11}$.
This situation persists in the latest PDG compilation \cite{PDG},
where $N(1650)S_{11}$ is now assigned a small but non-zero $\eta$ width.
The results of our analysis suggest that the large $\eta$ width for the
$N(1535)S_{11}$ is not due to a conventional $q^3$ state. One possible
explanation is the presence of another state in the same mass region,
{\it e.g.} a quasi-bound meson-baryon $S$ wave resonance just below
or above threshold, for example $N\eta$, $K\Sigma$ or $K\Lambda$
\cite{Kaiser}. Another possibility is an exotic configuration of four
quarks and one antiquark ($q^{4}\bar{q}$).

\section{Summary and conclusions}
\setcounter{equation}{0}

In this article, we have exploited the algebraic approach to baryon
structure to analyze simultaneously elastic
form factors and helicity amplitudes in photo- and electroproduction
and strong decay widths.
The use of algebraic methods allows us to study different situations,
such as the harmonic oscillator quark model and the collective model,
within the same framework. The logic of the
method is that, by starting from the charge and magnetization distribution
of the ground state (assuming a dipole form to the elastic form factor of
the nucleon), one can obtain the transition form factors to the excited
states. In the `collective' model, this procedure
yields {\it a power dependence of all form factors} (elastic
and inelastic) on $Q^2$.
For electromagnetic couplings we
have analyzed two aspects of hadronic structure: (i) the
breaking of $SU_{sf}(6)$ symmetry, 
and (ii) the stretching of hadrons
with increasing excitation energy. We find that, whereas the breaking
of the spin-flavor symmetry hardly effects the helicity amplitudes,
the stretching of hadrons does have a noticeable influence.

The disagreement between experimental and theoretical elastic form
factors and helicity amplitudes in the low-$Q^2$ region
$0\leq Q^2\leq 1$ (GeV/c)$^2$ may be due to coupling of the photon
to the meson cloud, ({\it i.e.} configurations of the type
$q^3-q\bar{q}$). Since such configurations
have much larger spatial extent than $q^3$, their effects
are expected to drop faster with momentum
transfer $Q^2$ than the constituent form factors.
Also, since meson exchange corrections contribute differently to different
channels, this effect will be state dependent.

We have performed a calculation of the strong decay widths
$N^{\ast} \rightarrow N + \pi$,
$N^{\ast} \rightarrow \Delta + \pi$,
$N^{\ast} \rightarrow N + \eta$,
$\Delta^{\ast} \rightarrow N + \pi$,
$\Delta^{\ast} \rightarrow \Delta + \pi$ and
$\Delta^{\ast} \rightarrow \Delta + \eta$ in a collective model of baryons.
The analysis of experimental data shows that,
while the decays into $\pi$ follow the expected pattern, the decays
into $\eta$ have some unusual features.
Our calculations do not show any indication for a large $\eta$ width,
as is observed for the $N(1535)S_{11}$ resonance. The observed large
$\eta$ width indicates the presence of another configuration, which is
outside the present model space. This suggests, that
in order to elucidate this point, particular attention be paid at
CEBAF to the $N \eta$ channel.

In the present contribution we have focused the discussion to the
nonstrange sector (nucleon and delta). It will be of interest to
extend the formalism to include strange baryons as well.

\section*{Acknowledgements}

The results reported in these proceedings are based on work done
in collaboration with F. Iachello (Yale).
The work is supported in part by grant No. 94-00059 from the United 
States-Israel Binational Science Foundation (BSF), Jerusalem, Israel 
(A.L.) and by CONACyT, M\'exico under project 
400340-5-3401E and DGAPA-UNAM under project IN105194 (R.B.).

\clearpage

\section*{Figure Captions}
\begin{description}
\item[Fig. 1] 
Collective model of baryons (the charge distribution of the 
proton is shown as an example.
\item[Fig. 2] 
Nucleon and delta mass spectrum ($M$ {\it vs.} $J^{P}$). 
Collective model \cite{BIL} $(+)$, 
nonrelativistic quark model \cite{NRQM} $(\times)$, relativized 
quark model \cite{RQM} $(\diamond)$.
\item[Fig. 3] 
Probability distribution of a typical ground state collective
oblate-top ground state wave function in an harmonic oscillator basis.
\item[Fig. 4]
Comparison between the experimental proton ($G_E^p$) and 
neutron ($G_E^n$) electric form factors with
the corresponding collective form factors. Dashed (solid) lines 
correspond to a calculation with (without) flavor breaking.
The form factors are divided by the dipole form
factor, $F_D=1/(1+Q^2/0.71)^2$.
\item[Fig. 5]
Comparison between the experimental proton ($G_M^p/\mu_p$) and 
neutron ($G_M^n/\mu_n$) magnetic form factors with
the corresponding collective form factors. Dashed (solid) lines 
correspond to a calculation with (without) flavor breaking.
The form factors are divided by the dipole form
factor, $F_D=1/(1+Q^2/0.71)^2$.
\item[Fig. 6]
Proton helicity amplitudes for excitation of $N(1520)D_{13}$
(a factor of $+i$ is suppressed) and $N(1680)F_{15}$.
The calculation with and without flavor breaking are shown by dashed and
solid lines, respectively.
\item[Fig. 7]
Effect of hadron swelling for excitation of
$N(1520)D_{13}$ (a factor of $+i$ is suppressed)
and $N(1680)F_{15}$.
The curves are labelled by the value of the stretching parameter $\xi$
of Eq.~(\ref{stretch}).
\item[Fig. 8]
Strong decay widths for $\Delta^{\ast} \rightarrow N + \pi$ and
$\Delta^{\ast} \rightarrow \Delta + \pi$ decays
of positive parity resonances with $L^P=2^+$ and negative parity
resonances with $L^P=1^-$.
The theoretical values are in parenthesis. All values in MeV.
\end{description}

\clearpage

\begin{figure}
\centerline{\hbox{
\psfig{figure=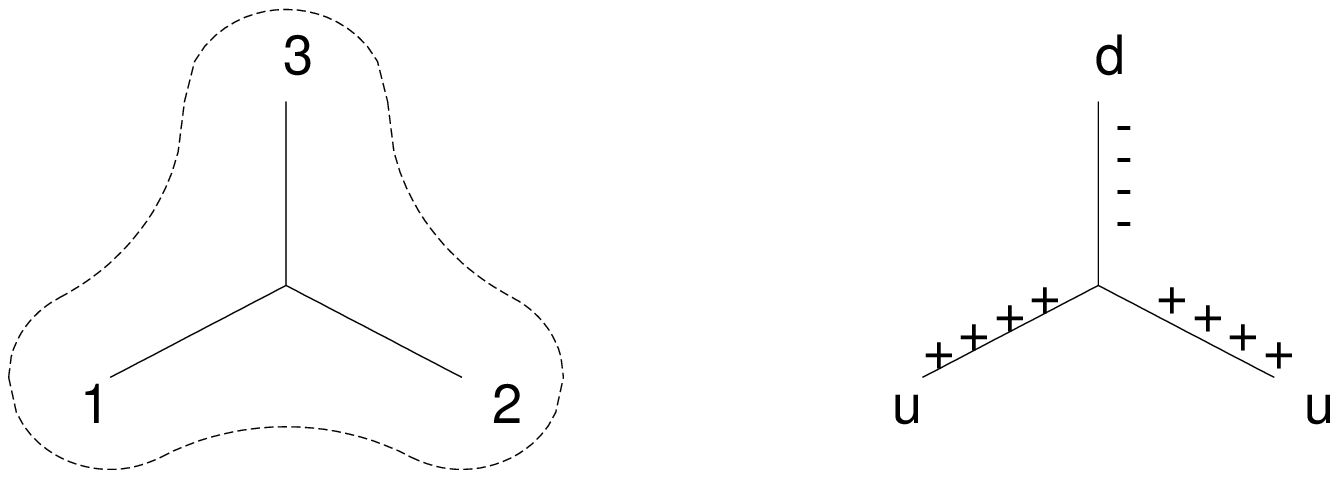} }}
\label{geometry}
\end{figure}

\clearpage

\begin{figure}
\centerline{\hbox{
\psfig{figure=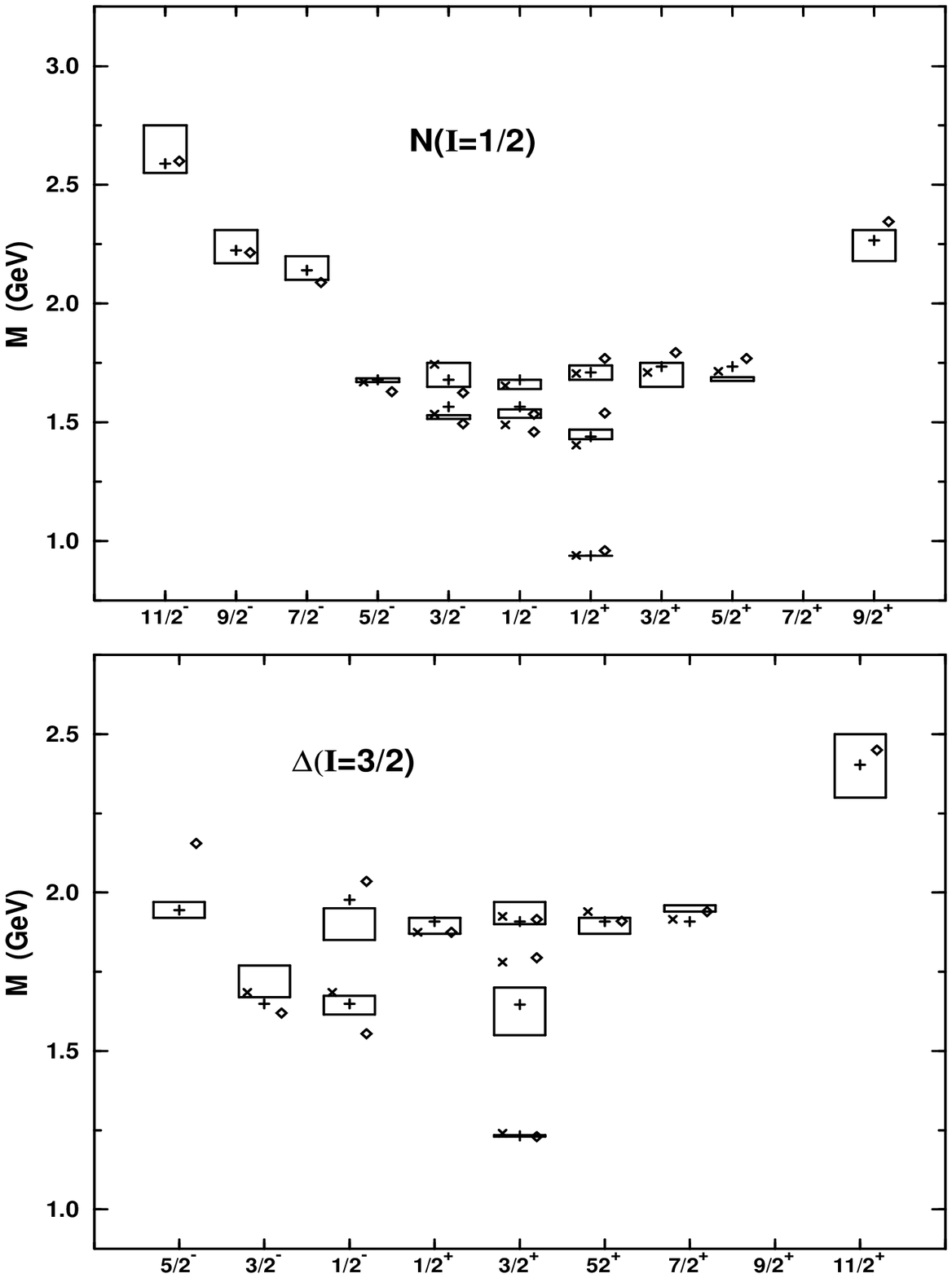} }}
\label{collef}
\end{figure}

\clearpage

\begin{figure}
\centerline{\hbox{
\psfig{figure=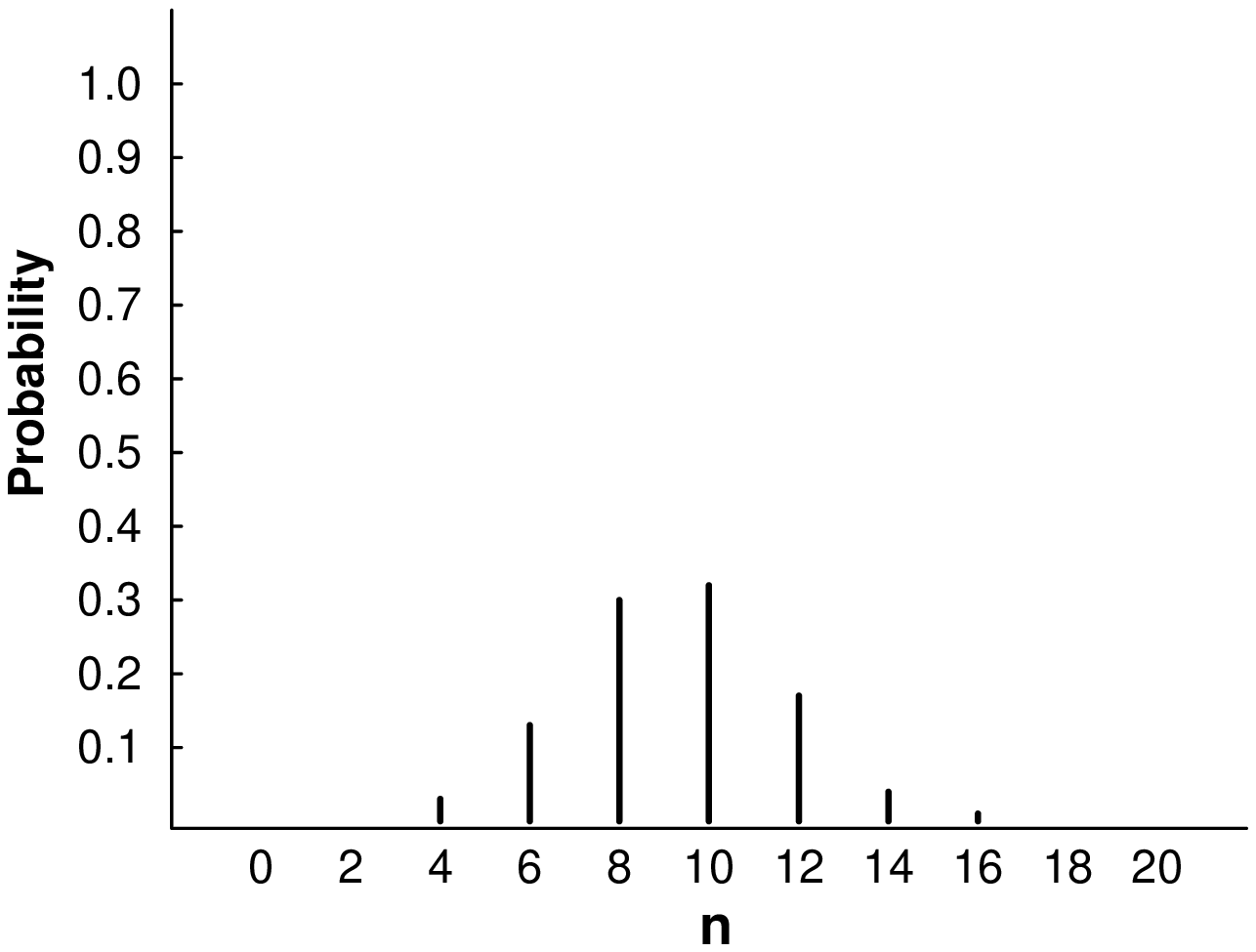} }}
\label{nuclmass}
\end{figure}

\clearpage

\begin{figure}
\centerline{\hbox{
\psfig{figure=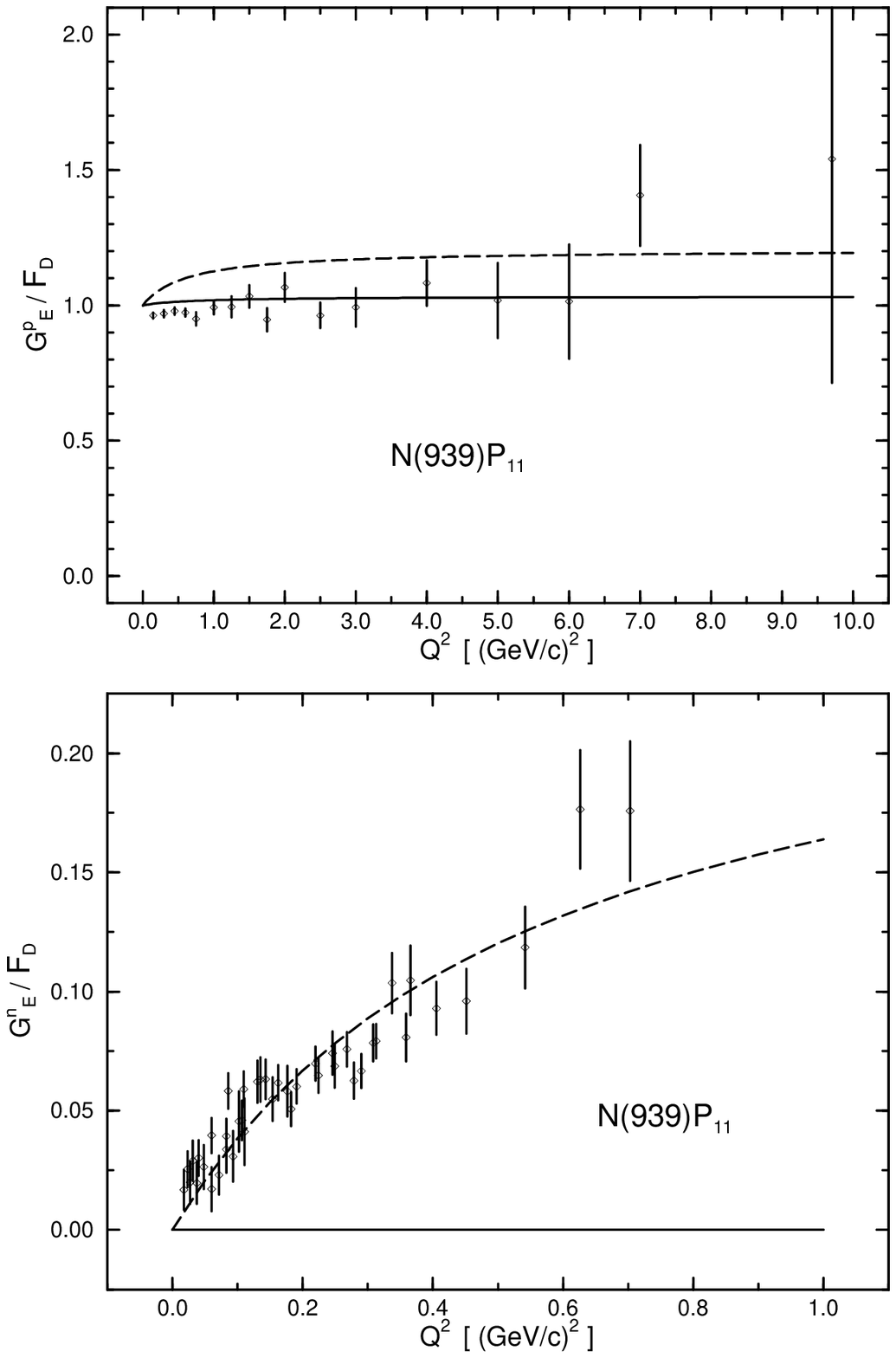} }}
\label{electricff}
\end{figure}

\clearpage

\begin{figure}
\centerline{\hbox{
\psfig{figure=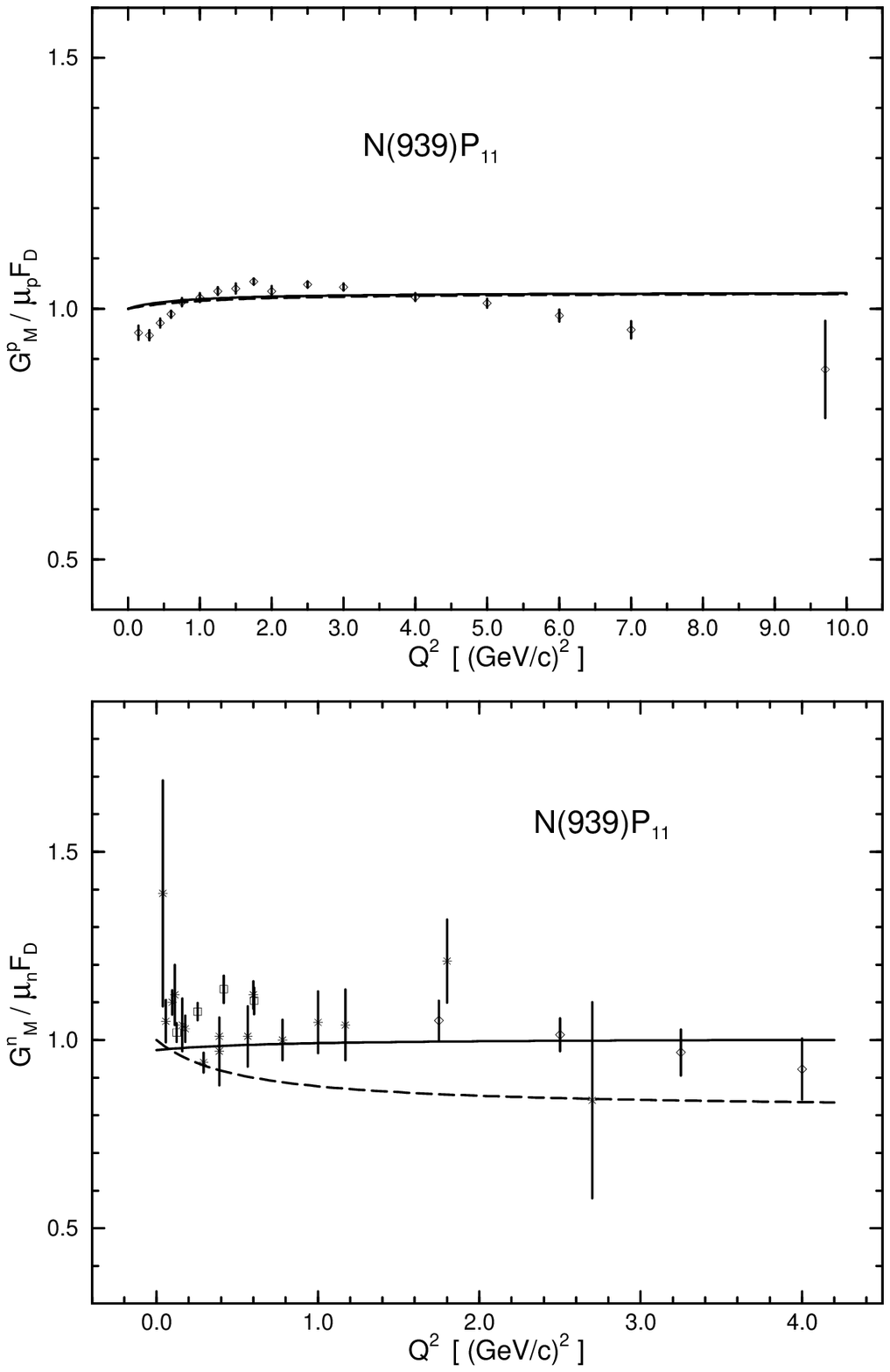} }}
\label{magneticff}
\end{figure}

\clearpage

\begin{figure}
\centerline{\hbox{
\psfig{figure=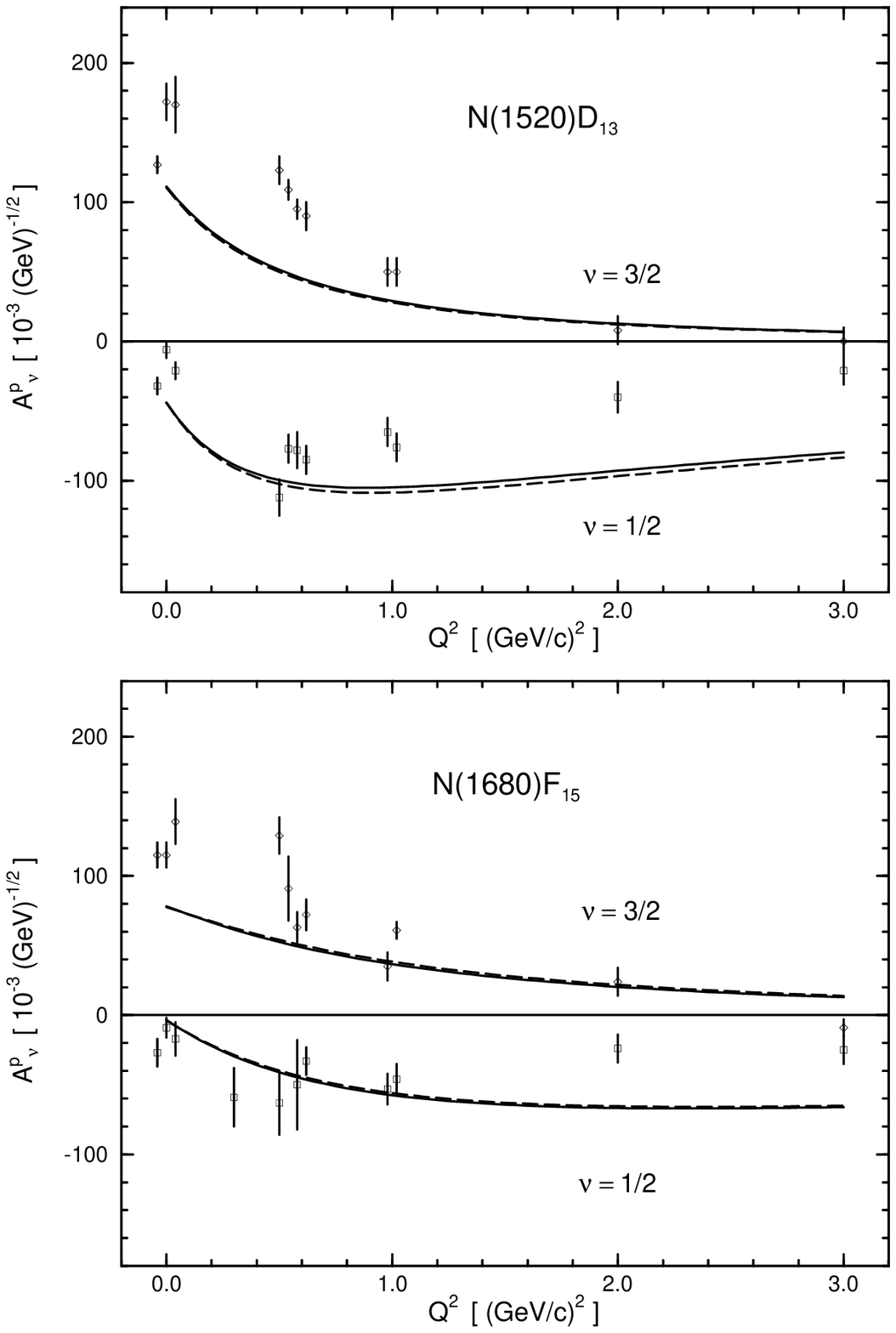} }}
\label{helicityampl}
\end{figure}

\begin{figure}
\centerline{\hbox{
\psfig{figure=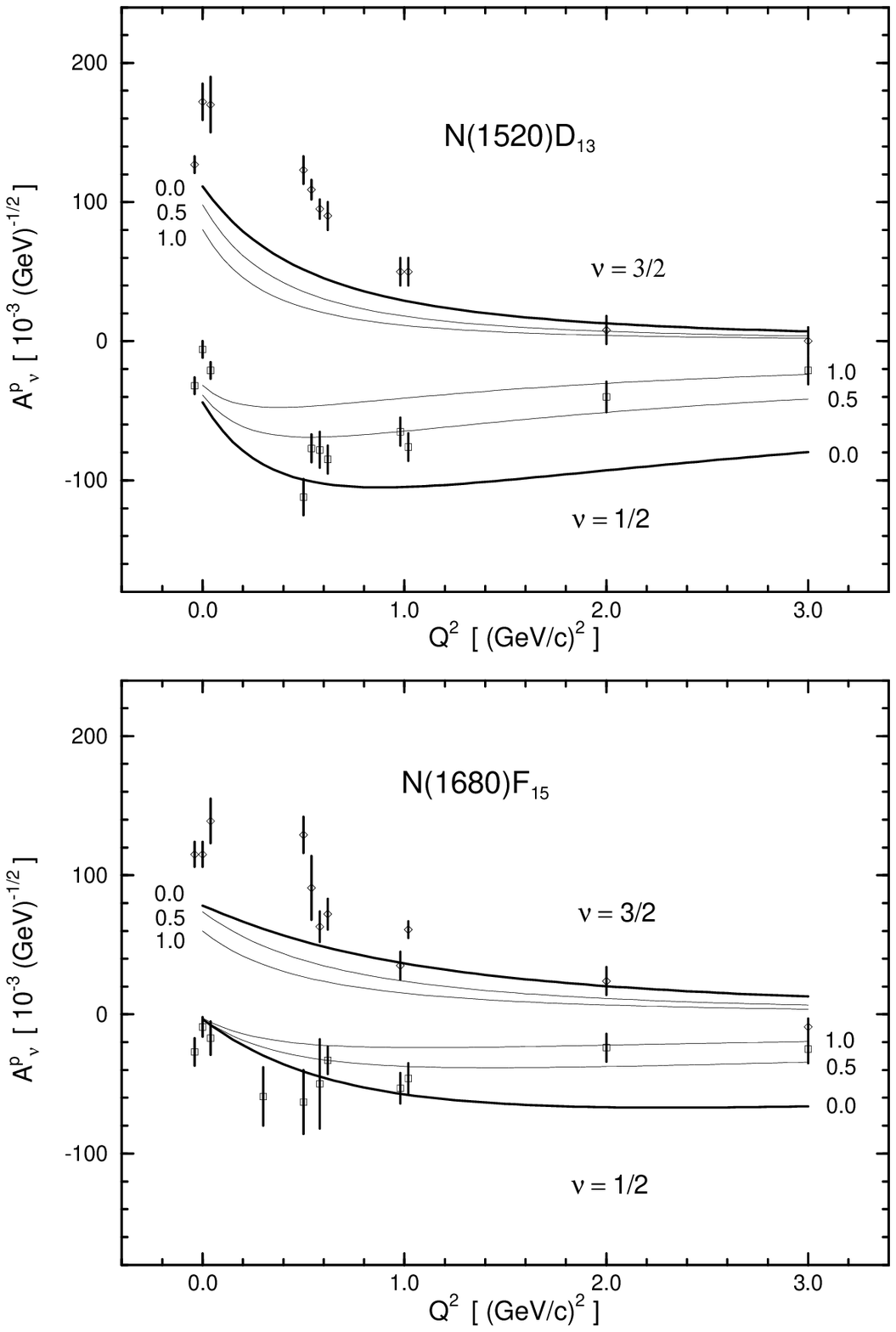} }}
\label{stretching}
\end{figure}

\clearpage
\begin{figure}
\centerline{\hbox{
\psfig{figure=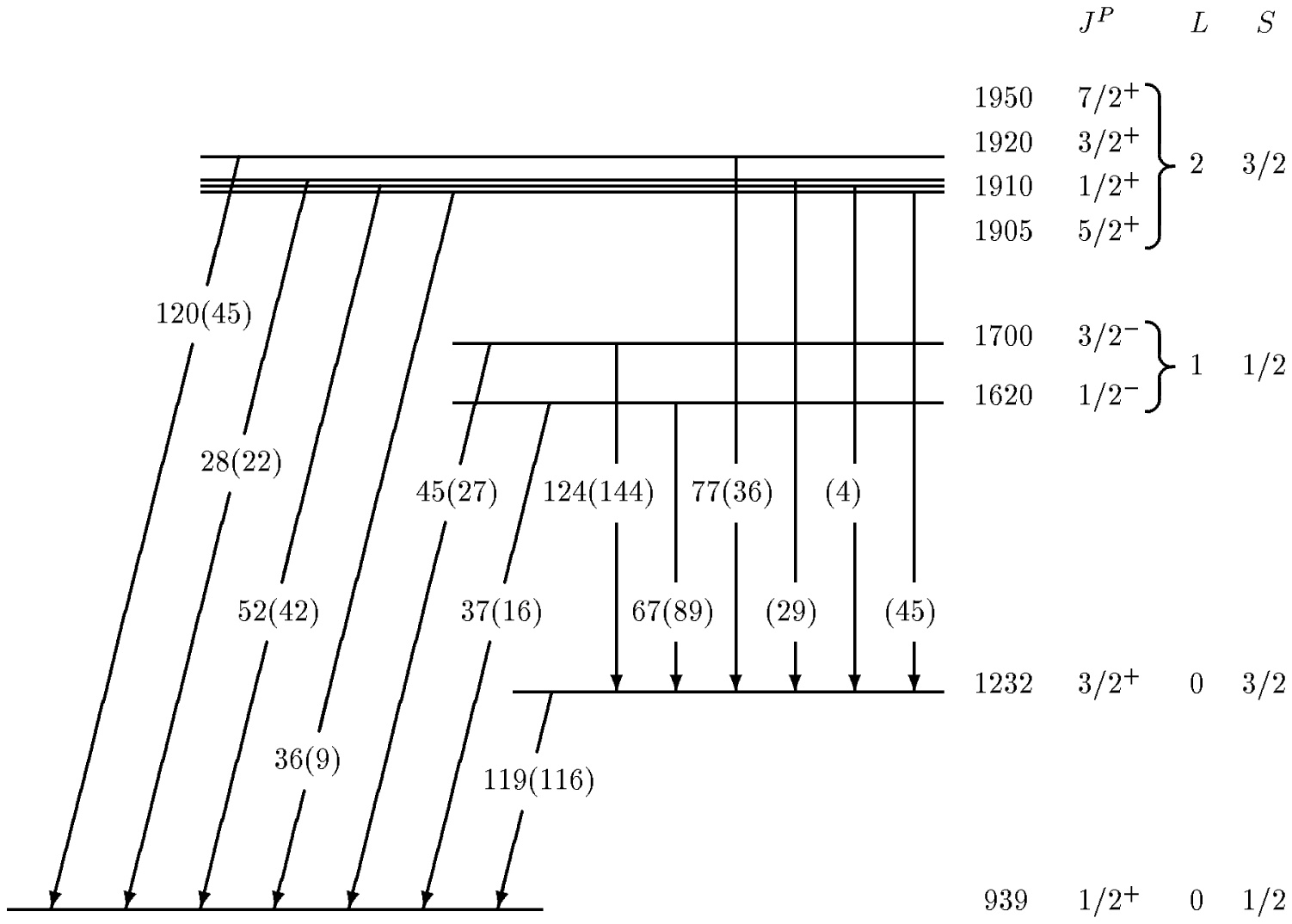} }}
\label{dwidths}
\end{figure}

\clearpage
\begin{table}
\centering
\caption[Collective and harmonic oscillator form factors]{\small
$a)$ Collective form factors (for large model space). The final states
are labeled by $[\mbox{dim}\{SU_{sf}(6)\},L^P]_{(v_1,v_2);K}$.
The initial state is $[56,0^{+}]_{(0,0);0}$.
$H(x)=\arctan x - x/(1+x^2)$.
$b)$ Harmonic oscillator form factors. The final states are
labeled by $[\mbox{dim}\{SU_{sf}(6)\},L^P]_{n}$.
The initial state is $[56,0^+]_0$.
\normalsize} \label{cff}
\vspace{15pt}
\begin{tabular}{cccc}
\hline
& & & \\
Final state$^{a)}$ & ${\cal F}(k)$
& ${\cal G}_{z}(k)/m_3 k_0 a$
& ${\cal G}_{\pm}(k)/m_3 k_0 a$ \\
& & & \\
\hline
& & & \\
$[56,0^+]_{(0,0);0}$
& $\frac{1}{(1+k^2a^2)^2}$ & $\frac{4ka}{(1+k^2a^2)^3}$ & 0  \\
& & & \\
$[70,1^-]_{(0,0);1}$ & $-i \, \sqrt{3} \, \frac{ka}{(1+k^2a^2)^2}$
& $i \, \sqrt{3} \, \frac{1-3k^2a^2}{(1+k^2a^2)^3}$
& $\mp i \, \sqrt{6} \, \frac{1}{(1+k^2a^2)^2}$ \\
& & & \\
$[56,2^+]_{(0,0);0}$
& $ \frac{1}{2} \sqrt{5}\left[ \frac{-1}{(1+k^2a^2)^2} \right.$
& $-\frac{1}{2} \sqrt{5}\left[ \frac{3+7k^2a^2}{ka(1+k^2a^2)^3} \right.$
& $\pm \sqrt{\frac{15}{2}}\left[ \frac{-1}{ka(1+k^2a^2)^2} \right.$ \\
& $\left. \hspace{1cm} + \frac{3}{2k^3a^3} H(ka) \right]$
& $\left. \hspace{1cm} - \frac{9}{2k^4a^4} H(ka) \right]$
& $\left. \hspace{1cm} + \frac{3}{2k^4a^4} H(ka) \right]$ \\
& & & \\
\hline
\hline
& & & \\
Final state$^{b)}$ & $\langle f | \hat U | i \rangle$
& $\langle f | \hat T_{z} | i \rangle/m_3 k_0 \beta$
& $\langle f | \hat T_{\pm} | i \rangle/m_3 k_0 \beta$ \\
& & & \\
\hline
& & & \\
$[56,0^+]_0$ & $\mbox{e}^{-k^2 \beta^2/6}$
& $\frac{1}{3} k \beta \, \mbox{e}^{-k^2 \beta^2/6}$ & 0 \\
& & & \\
$[70,1^-]_1$ & $-i \frac{1}{\sqrt{3}} k \beta
\, \mbox{e}^{-k^2 \beta^2/6}$
& $i \frac{1}{\sqrt{3}}(1-\frac{k^2 \beta^2}{3})
\, \mbox{e}^{-k^2 \beta^2/6}$
& $\mp i \sqrt{\frac{2}{3}} \, \mbox{e}^{-k^2 \beta^2/6}$ \\
& & & \\
$[56,2^+]_2$
& $-\frac{1}{3\sqrt{6}} k^2 \beta^2 \, \mbox{e}^{-k^2 \beta^2/6}$
& $\frac{2}{3\sqrt{6}} k \beta(1-\frac{k^2 \beta^2}{6})
\, \mbox{e}^{-k^2 \beta^2/6}$
& $\mp \frac{1}{3} k \beta \, \mbox{e}^{-k^2 \beta^2/6}$ \\
& & & \\
\hline
\end{tabular}
\end{table}

\clearpage
\begin{table}
\centering
\caption[]{\small
Strong decay widths for $\Delta^{\ast} \rightarrow N + \pi$
and $N^{\ast} \rightarrow N + \pi$ in MeV.
Experimental values are from \cite{PDG}.
\normalsize}
\label{regge}
\vspace{15pt}
\begin{tabular}{lcccccc}
\hline
& & & & & & \\
Resonance & $L$ & \multicolumn{4}{c} {$\Gamma$(th)} & $\Gamma$(exp) \\
& & Ref.~\cite{LeYaouanc} & Ref.~\cite{KI} & Ref.~\cite{CR} & Present & \\
& & & & & & \\
\hline
& & & & & & \\
$\Delta(1232)P_{33}$   & 0 & 70 & 121 & 108 & 116 & $119 \pm 5 $ \\
$\Delta(1950)F_{37}$   & 2 & 27 &  56 &  50 &  45 & $120 \pm 14$ \\
$\Delta(2420)H_{3,11}$ & 4 &  4 &     &   8 &  12 & $ 40 \pm 22$ \\
$\Delta(2950)K_{3,15}$ & 6 &  1 &     &   3 &   5 & $ 13 \pm 8 $ \\
& & & & & & \\
\hline
& & & & & & \\
$N(1520)D_{13}$   & 1 & & 85 & 74 & 115 & $67 \pm 9 $ \\
$N(2190)G_{17}$   & 3 & &    & 48 &  34 & $67 \pm 27$ \\
$N(2600)I_{1,11}$ & 5 & &    & 11 &   9 & $49 \pm 20$ \\
& & & & & & \\
\hline
\end{tabular}
\end{table}

\end{document}